\documentclass[iop]{emulateapj}

\usepackage{times}

\newcommand\fig[1]{Fig.~\ref{#1}}

\shorttitle{702 Alauda}
\shortauthors{Rojo and Margot}

\begin{document}

\title{Mass and density of B-type asteroid (702) Alauda}

\author{P. Rojo}
\affil{Departamento de Astronomia\\Universidad de Chile\\Santiago, Chile}
\email{pato@das.uchile.cl}
\author{J. L. Margot}
\affil{Departments of Earth and Space Sciences, Physics and Astronomy\\University of California, Los Angeles\\Los Angeles, CA 90095}
\email{jlm@astro.ucla.edu}

\begin{abstract}
Observations with the adaptive optics system on the Very Large
Telescope reveal that outer main belt asteroid (702) Alauda has a
small satellite with primary to secondary diameter ratio of $\sim$56.
The secondary revolves around the primary in 4.9143 $\pm$ 0.007 days
at a distance of 1227 $\pm$ 24 km, yielding a total system mass of
(6.057 $\pm$ 0.36) $\times$ 10$^{18}$ kg. Combined with an IRAS size
measurement, our data yield a bulk density for this B-type asteroid of
1570 $\pm$ 500 kg~m$^{-3}$.
\end{abstract}


\section{Introduction}

The discovery of solar system binaries has a considerable impact on
key problems in planetary science, partly because the binaries allow
direct measurements of fundamental physical and chemical properties
that are otherwise only obtainable with spacecraft.  Characteristics
of the mutual orbit provide crucial information about asteroid bulk
properties and internal structure, such as mass, density, porosity,
and mechanical strength.  These measurements are used to establish
links between asteroids and meteorites, and to understand the
geological context of meteorites that can be studied in great detail
in the laboratory.  Density measurements can also constrain the
proportion of ice to rock in distant minor planets, which is a strong
indicator of the chemical environment at the time of
formation~\citep{luni93, john05}.  The current proportion of binary
systems and their configuration are strong tracers of the collisional
and dynamical evolution of small bodies in the solar system.  Binaries
therefore illuminate the conditions in the solar nebula and early
solar system, and they help us refine our understanding of planet
formation.  The very high science priority of characterizing binary
systems has been reviewed by~\citet{weid89, merl02, marg02n, burn02,
  noll08}.

The third largest asteroid and former planet (2) Pallas (a=2.772 AU,
e=0.231, i=34.8$^\circ$) belongs to the taxonomic class B, which is
characterized by a linear featureless spectrum with bluish to neutral
slope in the wavelength range (0.435-0.925~$\mu$m)~\citep{bus02tax}.
In the case of Pallas, 
a broad absorption band in the 3~$\mu$m region is
detected~\citep{lars83, jone90} suggestive of hydrothermal alteration
and reminiscent of carbonaceous chondrite material~\citep{sato97}.
The negative spectral slope of B-type asteroids is attributed by
\citet{yang09} to a broad absorption band near 1.0 $\mu$m.
Because B-type asteroids may represent relatively primitive material
with similarities to carbonaceous meteorites~\citep{lars83, sato97,
  yang09}, their characterization can further our understanding of the
primordial building blocks of planets.
Unfortunately, our knowledge of the density of B-type asteroids is
poor, in part because few measurements exist, and in part because mass
estimates for (2) Pallas are widely scattered as they rely on indirect
techniques (gravitational perturbations on other asteroids or Mars).
\citet{carr10} reported a density of 3400 $\pm$ 900 kg~m$^{-3}$ based
on a mass estimate derived from pertubations to the orbits of
asteroids~\citep{mich00} with 22\% uncertainties and their volume
estimate with 4\% uncertainties.
\citet{schm09} reported a density of 2400 $\pm$ 250 kg~m$^{-3}$ based
on a mass estimate derived from perturbations to the orbit of
Mars~\citep{kono06} with 2.7\% uncertainties and their volume estimate
with 10\% uncertainties.  However their report lists mass estimates
for Pallas that vary by 20\%, and a more extensive survey indicates
mass estimates that vary by 60\%~\citep{hilt02}.  Using the mass
estimates of \citet{hilt99} and \citet{mich00} and the volume of
\citep{schm09} yields densities for Pallas in the range of 3700 $\pm$
390 kg~m$^{-3}$ and 2800 $\pm$ 670 kg~m$^{-3}$, respectively.
Here we report on the discovery~\citep{aster:cbet1016} and tracking of
the first satellite to a B-type asteroid, (702) Alauda (a=3.191 AU,
e=0.022, i=20.6$^\circ$), which allows us to provide a direct mass
measurement and improved density estimates for this class of
asteroids.

(702) Alauda has been identified as the largest member of a highly
inclined dynamical family in the outer main belt of
asteroids~\citep{fogl04,gil06}.  It is reasonable to assume that the
creation of this binary system dates back to the impact responsible
for the formation of the family~\citep{mich01,durd04}.  Because tidal
evolution depends on the formation age of the binary and the
mechanical properties of the components, our observations can also be
used to provide a relationship between formation age and mechanical
strength, specifically shear modulus and tidal dissipation
factor~\citep{tayl10icarus}.

\section{Observations and Data Reduction}
Using a highly optimized observing strategy, we were able to observe a
total of 111 unique main belt and Trojan asteroids during three nights
in visitor mode at the Very Large Telescope (VLT).  We used the
adaptive optics (AO) instrument NaCo
\citep{instrument:naco1,instrument:naco2} that has an 1024x1024 pixel
InSb array detector, which is fed by an adaptive optics system
consisting on Shack-Hartmann sensor with a 14x14 lenslet array.
Information from the wavefront sensor is used to control both a
tip-tilt mirror and a deformable mirror.  We used NaCo in its S13
mode, resulting in a pixel scale of 13.72 mas.

The observing sequence for each target consisted of four 10-second
integrations in each of four different offset positions.  Each frame
was then flat-fielded, and subtracted in pairs to remove the sky,
before alignment and addition to form a composite image.  We also
obtained a bad-pixel mask from outlier pixels present in the averaged
flat-field, and used the mask to ignore those pixels in each of the
10-second frame.

The companion to (702) Alauda was identified in a 40-second composite
image in the Ks filter ($\lambda_c = 2.18 \mu$m, FWHM $ = 0.35 \mu$m).
We eliminated the possibility of one type of AO artifact (a
non-rotating waffle mode) by immediately performing an observing
sequence with the field of view rotated 50 degrees. 
The non-sidereal tracking of the telescope discarded fixed background
sources and most moving background sources.  Follow-up observations
that night and the next fully confirmed the binarity of the system.

\begin{figure}
\includegraphics[width=0.9\linewidth]{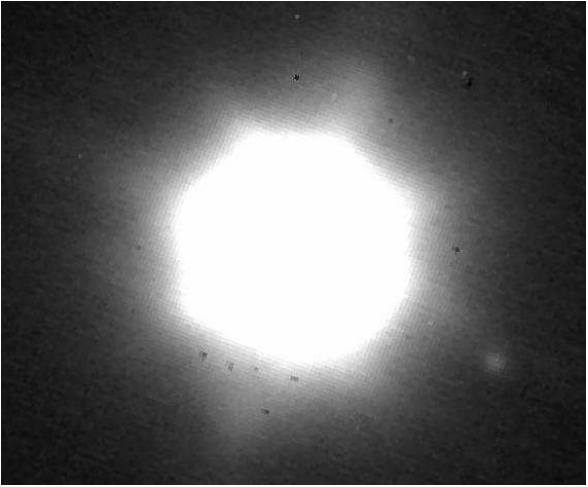}
\caption{\label{alauda}H-band image of the (702) Alauda binary system.
The companion is clearly seen in the lower-right quadrant.}
\end{figure}

The binary system was observed at 6 epochs during the last 2 nights of
the observing run with different filters.  The H-filter
($\lambda_c = 1.66 \mu$m, FWHM $ = 0.33 \mu$m) was found to give the best
compromise between SNR and AO correction (\fig{alauda}).  In order to
secure additional astrometric measurements, we submitted a Director's
Discretionary Time (DDT) proposal to follow-up the system in queue
mode, using the same observing instrument and strategy.  Table
\ref{tab-geom} summarizes all the observations of the binary system.

\begin{table*}
  \begin{center}
    \begin{tabular}{rrrrrrr}
UT Date and time & MJD   & Filter  &   RA (deg)  & DEC (deg)&  $\Delta$ (AU) &  Detection \\
\hline
2007 07 26 06:24 & 54307.2664 & Ks & 316.7047 &  -3.7504 &  2.1555 &  1    \\
2007 07 26 07:26 & 54307.3098 & H & 316.6957  &  -3.7485 &  2.1554 &  1    \\
2007 07 26 07:35 & 54307.3159 & J & 316.6942  &  -3.7482 &  2.1553 &  1    \\
2007 07 26 07:46 & 54307.3235 & H & 316.6927  &  -3.7479 &  2.1553 &  1    \\
2007 07 27 03:44 & 54308.1553 & H & 316.5163  &  -3.7124 &  2.1528 &  1    \\
2007 07 27 05:58 & 54308.2486 & H & 316.4965  &  -3.7085 &  2.1526 &  1    \\
\hline                                                                    
2007 08 18 04:07 & 54330.1716 & H & 311.7363  &  -3.2049 &  2.1565 &  1    \\
2007 08 29 04:12 & 54341.1750 & H & 309.6719  &  -3.2051 &  2.2082 &  0    \\
2007 09 09 01:45 & 54352.0731 & H & 308.1241  &  -3.2997 &  2.2884 &  1    \\
2007 09 10 02:45 & 54353.1147 & H & 308.0073  &  -3.3115 &  2.2974 &  0    \\
2007 09 11 01:39 & 54354.0685 & H & 307.9053  &  -3.3225 &  2.3059 &  1    \\
2007 09 12 04:09 & 54355.1727 & H & 307.7935  &  -3.3355 &  2.3159 &  0    \\
2007 09 13 02:12 & 54356.0920 & H & 307.7056  &  -3.3464 &  2.3245 &  0    \\
2007 09 14 01:46 & 54357.0738 & H & 307.6167  &  -3.3581 &  2.3338 &  1    \\
2007 09 15 01:10 & 54358.0488 & H & 307.5338  &  -3.3698 &  2.3432 &  0    \\
2007 09 16 02:18 & 54359.0959 & H & 307.4507  &  -3.3823 &  2.3535 &  0    \\
2007 09 20 01:06 & 54363.0457 & H & 307.1929  &  -3.4286 &  2.3940 &  0    \\
2007 09 21 00:54 & 54364.0377 & H & 307.1420  &  -3.4398 &  2.4046 &  0    \\
2007 09 28 01:56 & 54371.0804 & H & 306.9412  &  -3.5097 &  2.4841 &  0    \\
2007 09 30 03:32 & 54373.1471 & H & 306.9351  &  -3.5261 &  2.5086 &  0    \\
\hline
    \end{tabular}
    \caption{Observational circumstances.  Times listed refer to the
      start of the third exposure, roughly the middle of the
      four-image exposure sequence.  Observation epochs are given in
      Universal Time (UT) and Modified Julian Date (MJD) formats.  The
      Right Ascension (RA), Declination (DEC) and geocentric distance
      ($\Delta$) are used in orbital fits.  Observations above and
      below the line were obtained in visitor and queue mode,
      respectively.}
    \label{tab-geom}
  \end{center}
\end{table*}

We attempted to detect the companion in our composite images, and also
in images where an azimuthally averaged profile was subtracted from the
primary source.  We obtained ten secure detections.  The secondary may
be undetectable in other images due to proximity to the primary,
atmospheric seeing, lunar phase, or a combination of factors.  We
performed aperture photometry of the primary on the composite images,
and of the secondary on the profile-subtracted images, removing a
slanted-plane sky background where necessary.  We then combined the
primary and secondary photometry to measure primary-to-secondary flux
ratios and relative positions.

Our orbit determination software uses the separation and position
angle of the secondary with respect to the primary to solve for the
orbital parameters in the two-body problem.  The mutual orbital plane
orientation is assumed constant, but heliocentric motions and
light-time corrections are taken into account.  Astrometric positions
and their errors are specified at the mid-time of the exposure
sequence.  Starting from thousands of initial conditions, the software
adjusts for seven parameters (equivalent to six orbital elements plus
the mass of the system) in the nonlinear ordinary least squares
problem with a Levenberg-Marquardt technique.  The covariance matrix
and post-fit residuals are computed and inspected.  Binary orbits have
been computed with this algorithm for near-Earth
asteroids~\citep{marg02s,ostr06}, main-belt asteroids~\citep{marg03s,
merl02}, Kuiper belt objects~\citep{noll08,peti08}, dwarf planets, and
binary stars.

\section{Results}
\begin{figure}
\includegraphics[width=0.95\linewidth]{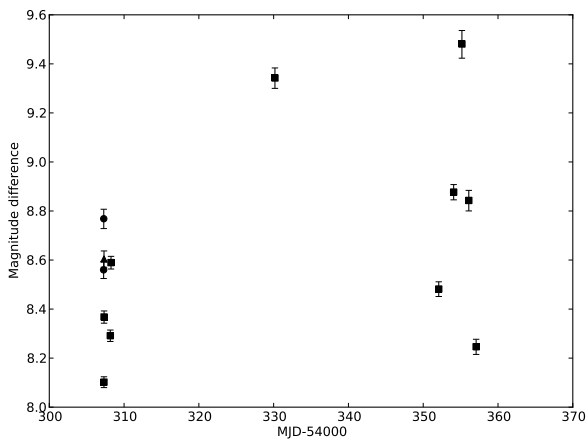}
\caption{\label{apphot}Primary-to-secondary magnitude differences for
  positive detections. Triangle, squares, and circles indicate the
  measurements using the J, H, and Ks filters, respectively.  The
  error bars indicate the Poisson uncertainty.}
\end{figure}

Primary-to-secondary flux ratios from the photometric analysis are
shown in \fig{apphot}.  Some, but not all, of the observed variations
may be related to the rotation of the primary with period
$8.3539\pm0.0007$ hrs and lightcurve amplitude of $0.09\pm0.02$
mag~\citep{beni08} or to the rotation of the secondary, which we
expect to be spin-locked (tidal despinning timescale $\sim$10$^6$
years).  Considering the theoretical Poisson noise ($\lesssim
0.05$mag), the differences in the same-night ratio measurements for
the J and H filters (0.21 and 0.27 magnitudes, respectively) must be
attributed to unknown systematics.
The average of the primary-to-secondary flux ratio is $3188 \pm 1444$
(magnitude difference of $8.8^{+0.4}_{-0.7}$), where the quoted
uncertainty is the standard deviation of our measurements.  This ratio
is one of the largest for solar system binaries, although a satellite
of (41) Daphne is reported to be 10 magnitudes fainter than the
primary~\citep{conr08}.  If we assume that the primary and secondary
have similar albedoes, the primary-to-secondary diameter ratio is
$\sim$56, and the mass of the secondary represents an insignificant
fraction of the total mass budget.

Relative positions of the components at the epochs of our measurements
are shown in Table \ref{tab-obs}. Since we only have one measurement per
epoch, as an upper limit we assigned centering uncertainties of 0.013
arcseconds in x and y, equivalent to the plate scale of the detector.

\begin{table*}[p]
  \begin{center}
    \begin{tabular}{rrrrrr} 
UT Date and time & MJD         & Sep (asec)&   PA (deg) & Res. (North) & Res. (East) \\
\hline
2007 07 26 06:24 & 54307.2663  & 0.584  &   21.3  &   0.975 &  0.588 \\  
2007 07 26 07:46 & 54307.3235  & 0.629  &   24.9  &  -0.081 & -0.167 \\
2007 07 27 03:44 & 54308.1553  & 0.751  &   58.0  &   0.942 & -0.446 \\
2007 07 27 05:58 & 54308.2486  & 0.726  &   60.5  &  -0.063 &  0.014 \\
\hline                                            							   
2007 08 18 04:07 & 54330.1716  & 0.748  &  232.9  &  -0.418 & -1.165 \\  
2007 09 09 01:45 & 54352.0731  & 0.751  &   46.4  &  -0.924 & -0.554 \\  
2007 09 11 01:39 & 54354.0685  & 0.651  &  213.4  &   0.530 &  0.357 \\  
2007 09 14 01:46 & 54357.0738  & 0.724  &   48.7  &  -0.640 & -0.230 \\  
\hline
    \end{tabular}
    \caption{Astrometry used in our orbital solution, showing the
    positions of the secondary relative to the primary, expressed as
    separations (Sep) in arcseconds and position angles (PA) in
    degrees East of North.  Epochs are given as in
    Table~\ref{tab-geom}.  Also shown are residuals from our best-fit solution
    in the format (O-C)/$\sigma$.}
    \label{tab-obs}
  \end{center}
\end{table*}

\begin{figure}
\includegraphics[angle=90, width=0.9\linewidth]{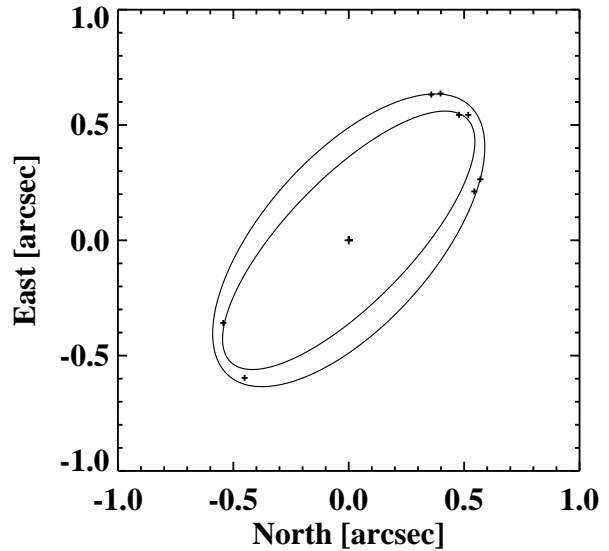}
\caption{\label{orbit} Observed secondary-to-primary separations and
  position angles, with error bars (crosses), and best-fit orbit
  (solid lines).  The exterior ellipse shows the projection of the
  orbit on the plane of the sky (62 degree inclination) on MJD 54307;
  it is representative of the first four astrometry data points at
  position angles 21, 25, 58, and 60 degrees.  The interior ellipse
  shows the projection of the best-fit orbit on the plane of the sky
  (69 degree inclination) fifty days later; it is more representative
  of the last three astrometry data points at position angles 46, 49
  and 213 degrees.}
\end{figure}

Our 7-parameter orbital fits (\fig{orbit}) used 16 measurements at 8 epochs.
The best-fit reduced chi-square value was 0.665, indicating that our
uncertainties were assigned somewhat conservatively but not overly so.
Values of the fitted parameters and their formal standard errors are
shown in Table~\ref{tab-fit}.  We measure a system mass of (6.057
$\pm$ 0.36) $\times$ 10$^{18}$ kg (roughly 10$^{-6}$ Earth masses),
corresponding to a fractional precision of 6\%, considerably better
than results from indirect techniques.

\begin{table}
\begin{center}
\begin{tabular}{llrr}
Parameter & & Estimate & $\sigma_{\rm formal}$\\
\hline
Orbital period [days]             & $P$      &     4.9143  &  0.007   \\    
Semi-major axis [km]              & $a$      &     1226.5  &  24      \\    
Eccentricity                      & $e$      &    0.003    &  0.02    \\
Longitude of ascending node [deg] & $\Omega$ &     280.9   &  3       \\
Inclination [deg]                 & $i$      &      46.5   &  3       \\ 
Argument of pericenter [deg]      & $\omega$ &    169.074  &  180     \\
Epoch of pericenter passage [MJD] & $T$      &  54308.648  &  4       \\    
\hline
\end{tabular}
\caption{Results of orbital fits to our astrometry, giving parameter
estimates and formal standard errors.  Angles are provided in the
equatorial frame of epoch J2000.  The orbit normal is at right
ascension 190.9$^\circ$ and declination 43.5$^\circ$, corresponding to
ecliptic longitude 168.3$^\circ$ and latitude 43.3$^\circ$.  The
orientation of pericenter and time of pericenter passage are poorly
constrained, as expected for such a low eccentricity orbit.
Nevertheless we provide guard digits in those quantities to allow
computation of the relative positions of the components without loss
of precision.}
\label{tab-fit}
\end{center}
\end{table}

The IRAS Minor Planet Survey (IMPS) lists the diameter of (702) Alauda
as 194.73 $\pm$ 3.2~km.  Combining our mass measurement with this size
estimate, we compute a density with formal uncertainties of
1570 $\pm$ 120 kg~m$^{-3}$.  Realistic uncertainties need to take into
account the fact that there may be a bias in the asteroid size and
that the shape of the asteroid may differ substantially from a sphere.
The IMPS diameter measurements have been compared with occultation
results and are believed to be accurate to 10\%~\citep{tede02}.
Therefore we suggest a density estimate with more realistic errors of
1570 $\pm$ 500 kg~m$^{-3}$.

\begin{figure}[htbp]
  \begin{center}
     \includegraphics[angle=90,width=0.9\linewidth]{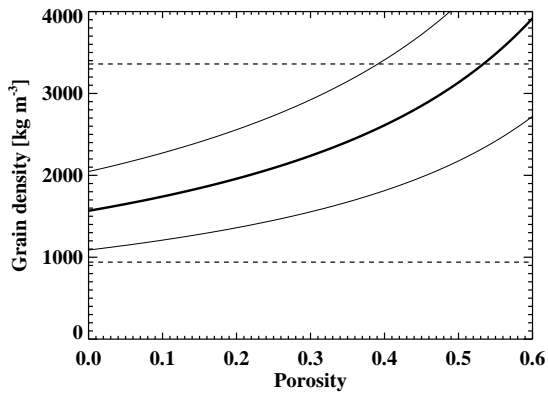} 
     \caption{Range of porosities and grain densities admitted by our
       density measurement (our nominal solution is the bold solid
       line, with density uncertainties captured by the solid lines
       above and below). The densities of water ice and anhydrous
       silicates are shown as dashed lines.}
    \label{fig-porosity}
  \end{center}
\end{figure}

The confounding effects of porosity must be kept in mind in
interpreting density measurements~\citep{brit02,mcki08}. In
Fig.~\ref{fig-porosity} we show the range of grain densities that are
compatible with our bulk density for several values of porosity.
Because the pressure at the center of (702) Alauda is moderate
($\sim$3.3 MPa or $\sim$33 atmospheres), this body could in principle
sustain fairly large porosities.  At near-zero porosities a mix of
anhydrous silicates and water ice would require a fair amount of ice
(roughly 3/4 by volume or 4/9 by mass) to match the density
constraint.  At high porosities the asteroid could be entirely devoid
of ice.
Progress in fully understanding the manifestation of porosity in small
bodies will require a large sample of reliable sizes and densities, or
in-situ seismic experiments.

\pagebreak
\section{Conclusions}
We report the discovery of a satellite to (702) Alauda, a member of an
asteroid collisional family.  The satellite is small, with a ratio of
primary to secondary radii of about 50.  The secondary revolves around
the primary in 4.91 days at a distance of 1,230 km, corresponding to
12.6 primary radii.  Due to tidal interactions, its spin period is
expected to synchronize to its orbital period on million year
timescales.  The secondary likely formed by a sub-catastrophic
collision on the primary, and subsequently tidally evolved outward
\citep{merl02,durd04}.  Because the orbit is quasi-circular it is
likely that the secondary is mechanically weaker than the
primary~\citep{marg03s,tayl10icarus}, as might be expected if the body
formed by the re-accumulation of impact ejecta.  Our mass measurement
with 6\% uncertainties provides an important new constraint on the
composition and internal structure of B-type asteroids.  Combined with
an IRAS size measurement, the mass yields a bulk density of 1570 $\pm$
500 kg~m$^{-3}$.  This density admits an ice-to-rock mass ratio in the
range 0-80\%, depending on the porosity of the asteroid.  Our density
determination is significantly lower than previous estimates for the
B-type asteroid (2) Pallas, which may be related to different
porosities.

\acknowledgments We thank referee W.\ Merline for insightful comments
and F. Selman, C. Lidman, and N. Ageorges at the VLT for assistance
with the observations.  We are also grateful for the allocation of
telescope time for follow-up observations, without which this binary
could not have been fully characterized.  PMR was supported by Center
of Excellence in Astrophysics and Associated Technologies (PFB 06),
FONDAP 15010003, and Fondecyt grant 11080271. JLM was supported by
NASA Planetary Astronomy grant NNX07AK68G/NNX09AQ68G.

\clearpage 

\bibliography{ms}

\end{document}